\documentclass{article}

\usepackage{arxiv}

\usepackage[utf8]{inputenc} 
\usepackage[T1]{fontenc}    
\usepackage{hyperref}       
\usepackage{url}            
\usepackage{booktabs}       
\usepackage{amsfonts}       
\usepackage{nicefrac}       
\usepackage{microtype}      
\usepackage{lipsum}		
\usepackage{graphicx}
\usepackage{natbib}
\usepackage{doi}

\usepackage{mathtools} 

\usepackage[autostyle=true]{csquotes}
\usepackage{tikz}
\usepackage[normalem]{ulem}
\usetikzlibrary{calc}
\usetikzlibrary{3d}

\title{Optimal design of photonic nanojets under uncertainty}

\author{Amal Mohammed A Alghamdi\\
	Department of Applied Mathematics and Computer Science\\
	Technical University of Denmark\\
	DK-2800 Kgs. Lyngby, Denmark 
	\And
	Peng Chen \\
	School of Computational
Science and Engineering\\
	Georgia Institute of Technology\\
	Atlanta, GA 30308, USA \\
	\And
	Mirza Karamehmedovi\'c \\
	Department of Applied Mathematics and Computer Science\\
	Technical University of Denmark\\
	DK-2800 Kgs. Lyngby, Denmark 
}

\begin{document}
\maketitle

\begin{abstract}
Photonic nanojets (PNJs) have promising applications as optical probes in super-resolution optical microscopy, Raman microscopy, as well as fluorescence microscopy. In this work, we consider optimal design of PNJs using a heterogeneous lens refractive index with a fixed lens geometry and uniform plane wave illumination. In particular, we consider the presence of manufacturing error of heterogeneous lens, and propose a computational framework of Optimization Under Uncertainty (OUU) for robust optimal design of PNJ. We formulate a risk-averse stochastic optimization problem with the objective to minimize both the mean and the variance of a target function, which is constrained by the Helmholtz equation that governs the 2D transverse electric (2D TE) electromagnetic field in a neighborhood of the lens. The design variable is taken as a spatially-varying field variable, where we use a finite element method for its discretization, impose a total variation penalty to promote its sparsity, and employ an adjoint-based BFGS method to solve the resulting high-dimensional optimization problem. We demonstrate that our proposed OUU computational framework can achieve more robust optimal design than a deterministic optimization scheme to significantly mitigate the impact of manufacturing uncertainty.
\end{abstract}


\section{Introduction}\label{sec:introduction}
Photonic nanojets (PNJs) are highly focused beams of light that can arise, e.g., when a laser illuminates a micrometer-sized glass lens. Illuminating a nano-sample with a PNJ produces a far field that is hyper-sensitive to the presence, the properties, and the position of the sample relative to the PNJ. For example~\cite{Chen:04,Li:05}, placing a 3D gold nanoparticle sized 2-60 nm inside a PNJ introduces a perturbation in the backscattered far-field intensity anywhere from -35 dB to 15 dB relative to the backscattering intensity of the PNJ-producing micro-lens alone. This
and other effects, like the resonant interaction of PNJs with dielectric and plasmonic nanostructures in
the sample, make PNJ illumination a promising prospect for simple and inexpensive superresolution optical detection, localization, measurement, imaging, and manipulation~\cite{darafsheh2021photonic,SPIE}.

There are many numerical and a few experimental works attempting PNJ design by structured illumination or by lens shaping~\cite{Lecler-2019,Zhu-2016,Paganini-2015}. Lens shaping optimizes the geometry of a homogeneous lens~\cite{9497753,Hengyu:15} for a fixed illumination and may employ additional structures such as planar supporting substrates, planar substrates with gaps, and combinations of two or more lenses. Some lens shaping studies have allowed limited heterogeneity in the lens refractive index (e.g. layered spheres)~\cite{Geints:11}. So far,
point-, line-, hollow-focus and multiple-foci PNJ were obtained by structured illumination of dielectric
microspheres, and PNJ position was designed by illuminating micro-scale dielectric spheres,
hemispheres and spherical caps, dielectric elliptical particles, cylinders/disks, cuboids, core-shell
spheres, core-shell cylinders and core-shell cuboids, spiral axicons, chains of metal-dielectric
spheres, chains of core-shell cylinders and low-dimensional parameterized dielectric particles. Recently, \cite{karamehmedovicEtAl2022} studied dynamic PNJ steering using computed structured illumination and a fixed simple homogeneous lens, while \cite{SPIE} applied this steerable optical probe in a numerical demonstration of lateral and vertical super-resolution optical imaging.

The PNJ design scheme we propose and study here involves a fixed-geometry heterogeneous micro-lens. Since a given desired PNJ position requires a specific and detailed optimized lens refractive index profile, it is interesting to perform uncertainty quantification of the PNJ design scheme given lens manufacturing errors and illumination imprecision. Quantifying and mitigating these uncertainties is essential to produce robust optical systems. Nevertheless, there is limited attention in the literature to the effect of these uncertainties and taking them into account in the optimization process.

Our goal with this work is twofold. First, we devise and demonstrate numerically a systematic approach to design PNJ properties, in particular, the PNJ location and intensity. We achieve this by optimizing the heterogeneous lens profile for fixed lens geometry and fixed uniform plane wave illumination. The lens heterogeneity provides flexibility in the PNJ design, as it increases the number of degrees of freedom (DOFs) in the optimal solution. This flexibility eliminates the need for using complex lens shapes and enables achieving the design using, e.g., cylindrical and spherical lenses. Next, we quantify and mitigate the effect of possible manufacturing error using an Optimization Under Uncertainty (OUU) framework. The numerical feasibility of our approach is realized using scalable high-dimensional optimization techniques for PDE-based problems.

We do not expect our computed lens profiles to be experimentally realizable at the moment. Rather, our work is a numerical analysis of the effects of manufacturing and illumination uncertainties on the performance of the proposed PNJ design. In the microwave regime, however, with wavelengths of the order of 1 cm, we do expect the lens profiles to be feasible to produce.

\begin{figure}[!htb]
    \centering
    \includegraphics[width=3.2in]{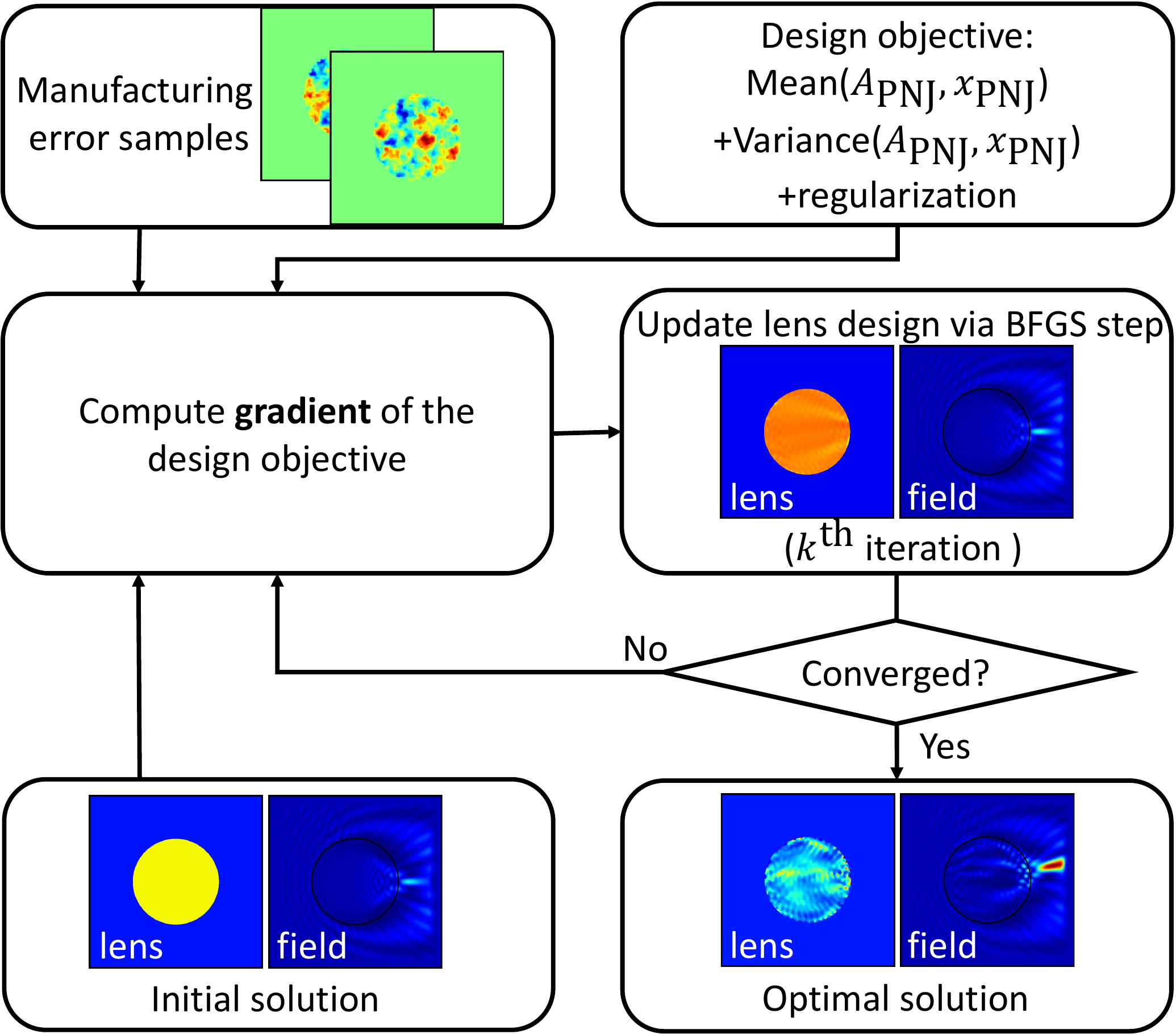}
    \caption{Conceptual illustration of the optimization under uncertainty (OUU) framework. $x_\text{PNJ}$ and $A_\text{PNJ}$ are the desired PNJ location and amplitude, respectively.}
    \label{fig:concept}
\end{figure}

Fig.~\ref{fig:concept} shows a conceptual illustration of the sample average approximation (SAA)-based OUU framework that we build in section \ref{sec:The Control Problem}. The framework needs as inputs a representation of the manufacturing noise and a design objective, Fig.~\ref{fig:concept} (top). The latter is normally written in terms of expected values of some stochastic objective and a regularization term. The optimization process starts with an initial design, Fig.~\ref{fig:concept}, (bottom left) at which the objective and its gradient with respect to the design parameters are computed. The gradient is used to create a BFGS step to update the initial design. The framework terminates if the updated design achieves the convergence criteria, otherwise it will continue to create the next BFGS step and repeat the process.  

\section{Lens optimization and PNJ design.}\label{sec:The Control Problem}

\subsection{PNJ Model}

\begin{figure}[!htb]
    \centering
    \includegraphics[width=1.6in]{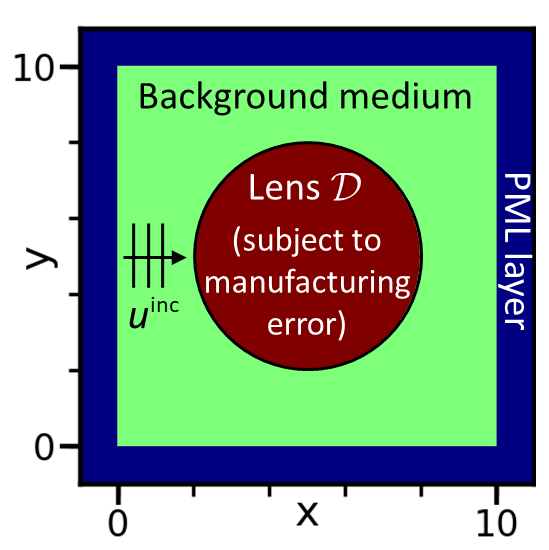}
    \caption{The setup of the forward scattering problem.}
    \label{fig:sketch}
\end{figure}

Fig.~\ref{fig:sketch} illustrates the finite element method (FEM) setup of our direct scattering problem. We consider the light-lens interaction in the time-harmonic 2D transverse electric (2DTE) case with the time-dependence factor $\exp(-i\omega t)$ suppressed. Thus the electric field vector always points outside the plane and may be represented by a complex scalar field $u$. Let $\lambda_0$ and $k_0=2\pi/\lambda_0$ be the operating free-space wavelength and wavenumber, respectively. When the lens is illuminated by the $\widehat{\textbf{x}}$-directed, normalized uniform plane wave $u^{\rm inc}(x,y)=\exp(ik_0x)$, the produced scattered field $u^{\rm sca}$ satisfies the Helmholtz system
\begin{align}
  & \Delta u^\text{sca} + k(x,y)^2u^\text{sca}=k_0^2(1 -n(x,y)^2)\exp(ik_0x) \quad \text{in}\; \mathbf{R}^2, \label{equ:helmholtz} \\
   &\lim_{r \rightarrow \infty} r^{\frac{1}{2}} \left(\frac{\partial u^\text{sca}}{\partial r} - i k u^\text{sca} \right) = 0\nonumber \\
    &\quad\quad\quad\quad \text{uniformly in all directions}\; \frac{(x,y)}{r}, \label{equ:summerfield_radiation}
\end{align}
where $r=\sqrt{x^2+y^2}$, $n(x,y)\ge1$ is the heterogeneous lens refractive index profile for $(x,y)\in\mathcal{D}$, and $n(x,y)=1$ outside the lens. Finally, \eqref{equ:summerfield_radiation} is the Sommerfeld radiation condition (approximately modeled by a perfectly matched layer, PML, \cite{turkel1998absorbing}).

When referring to a PNJ we shall mean a feature of the total resulting field $u^{\rm tot}=u^{\rm inc}+u^{\rm sca}$ near the lens.

\subsection{Deterministic Design Problem Formulation}

To obtain the desired PNJ, we choose a design objective that maximizes the total wave amplitude at a desired PNJ location, $x_\text{PNJ} \in  \mathbf{R}^2$. The objective function $\mathcal{Q}$ is given by  

\begin{align}
\mathcal{Q}(u^\text{tot}(\tau)) = \frac{1}{2}\int_{ \mathbf{R}^2} \delta_{x_\text{PNJ}}(x)\left( |\Re u^\text{tot}(\tau)|^2 + |\Im u^\text{tot}(\tau)|^2 - A_\text{PNJ}^2 \right)^2,
\end{align}
where  $\Re u^\text{tot}$ and $\Im u^\text{tot}$ are the real and imaginary parts of the total wave, respectively. We parameterize the wave number as $k = k_0 + e^{\tau}\chi_\mathcal{D}$, where $\tau$ is the design variable and $\chi_\mathcal{D}$ is a characteristic function with support on the lens $\mathcal{D}$. $\delta_{x_\text{PNJ}}(x)$ is the Dirac delta at $x_\text{PNJ}$. $A_\text{PNJ}$ is the desired PNJ amplitude which we normally set to an unattainable high value to achieve the maximum possible amplitude in practise.

\subsection{Stochastic Design Problem Formulation}

In optical systems, a stochastic manufacturing error might be present, in the lens properties for example. Here, we assume a lens manufacturing error $\zeta$ that we represent as a Gaussian random field with Mat\'ern covariance. The stochastic design objective $\mathcal{Q}$ is now given by

\begin{align}
\mathcal{Q}(u^\text{tot}(\tau, \zeta)) =& \nonumber \\
\frac{1}{2}\int_{ \mathbf{R}^2} \delta_{x_\text{PNJ}}(x)&\left( |\Re u^\text{tot}(\tau, \zeta)|^2 + |\Im u^\text{tot}(\tau, \zeta)|^2 - A_\text{PNJ}^2 \right)^2 \label{equ:Q stochastic},
\end{align}
where we update the parameterization of the wave number $k$ to take into account the manufacturing error $\zeta$, $k = k_0 + e^{\tau + \zeta}\chi_\mathcal{D}$.

\begin{figure}[!htb]
    \centering
    \includegraphics[width=1.6in]{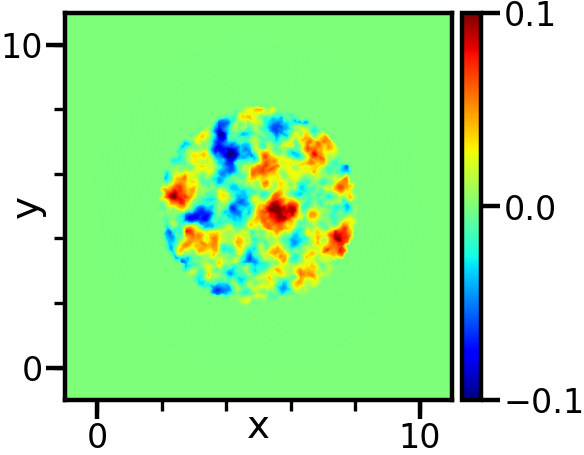}
    \includegraphics[width=1.6in]{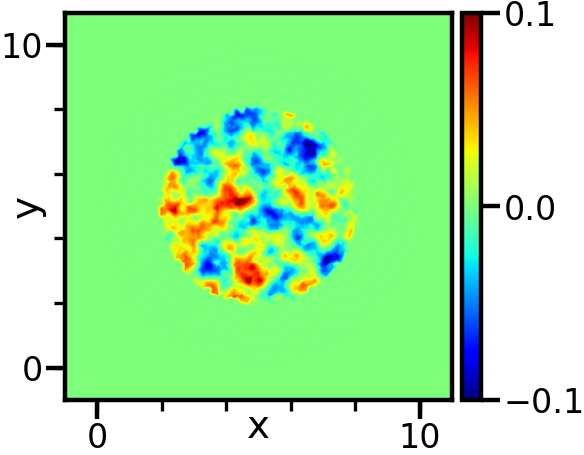}
    \caption{Two samples of the manufacturing error $\zeta \sim \mathcal{N}(0, \mathcal{C})$.}
    \label{fig:noise samples}
\end{figure}

We assume $\zeta \sim \mathcal{N}(0,\mathcal{C})$, where $\mathcal{N}$, is a zero-mean normal distribution and $\mathcal{C}$ is Mat\'ern-class covariance operator. A scalable approach to obtain samples from the  distribution $\mathcal{N}(0,\mathcal{C})$ is to employ the link between Mat\'ern-class Gaussian random fields and elliptic Stochastic Partial Differential Equations (SPDE) \cite{lindgren2010explicit}, see the appendix. In this approach, the covariance operator  $\mathcal{C}(\delta, \gamma, \alpha)$ is  parameterized by the scalar parameters $\delta$, $\gamma$ and $\alpha$ that collectively control the variance, smoothness, and correlation length of the field.  Fig.~\ref{fig:noise samples} shows samples of the described Gaussian random field for $\alpha=2$, $\gamma=2.5$, and $\delta=25$.

\subsection{Optimization under Manufacturing Uncertainty} \label{sec:The Control Problem: OUU}

To formulate the design objective \eqref{equ:Q stochastic} as an OUU problem, we use the risk-averse mean-variance formulation:
\begin{align}
\mathcal{J}(\tau) = \mathbb{E}_\zeta[\mathcal{Q}(\tau, \zeta)] + \beta_V \text{Var}_\zeta[\mathcal{Q}(\tau, \zeta)] + \beta_P P(\tau),
\label{equ:optimization functional}
\end{align}
with the goal to maximize the expectation of the total wave amplitude while minimizing its variability,
where $\mathbb{E}_\zeta$ and $ \text{Var}_\zeta$ denotes expected value and variance with respect to the manufacturing error $\zeta$, respectively. $\beta_V$ is a weight for the variance term. The last term in \eqref{equ:optimization functional} is a penalty (regularization) term, where $P(\tau) = \int_{\mathcal{D}} |\tau(x)| dx \approx  \int_{\mathcal{D}} (\tau^2(x)+\epsilon)^\frac{1}{2} dx $ is an approximate total variation penalty function that promotes sparsity of the material and $\beta_P$ is the penalty term weight. We tune $\beta_V$ and $\beta_P$ in practise to control the degree to which we enforce the corresponding terms.

\subsection{Numerical Methods}\label{sec:The Control Problem: numerical methods}
To evaluate the cost functional \eqref{equ:optimization functional} numerically, we use sample average approximation (SAA) where the mean and the variance in \eqref{equ:optimization functional} are approximated as follows
\begin{align}
\mathbb{E}_\zeta[\mathcal{Q}(\tau, \zeta)] &\approx \Bar{ \mathcal{Q}} \coloneqq \frac{1}{M} \sum_{m=1}^{M} \mathcal{Q}(\tau, \zeta_m), \\
\text{Var}_\zeta[\mathcal{Q}(\tau, \zeta)] &= \mathbb{E}_\zeta[\mathcal{Q}^2(\tau, \zeta)] - \mathbb{E}_\zeta[\mathcal{Q}(\tau, \zeta)]^2 \nonumber \\
&\approx  \frac{1}{M} \sum_{m=1}^{M} \mathcal{Q}^2(\tau, \zeta_m) - \Bar{\mathcal{Q}}^2, 
\end{align}
where $M$ is the number of samples used in the SAA.

In the objective \eqref{equ:Q stochastic}, we assume that the solution of the Helmholtz equation \eqref{equ:helmholtz}--\eqref{equ:summerfield_radiation}, the scattered field $ u^\text{sca}$, is independent from the design variable $\tau$. We enforce the dependency of the scattered field on $\tau$ via constraining the optimization problem with the weak form of the system \eqref{equ:helmholtz}--\eqref{equ:summerfield_radiation}, details of formulating the weak form for the Helmholtz equation can be found in \cite{chen2021optimal}. We denote the weak form of \eqref{equ:helmholtz}--\eqref{equ:summerfield_radiation} by
\begin{align}
a(u^\text{sca}_m, v_m;\tau, \zeta_m) = b(v_m) \quad \forall\;\text{test function}\;v_m \label{equ:weak form},
\end{align}
where $a$ and $b$ are the bilinear and linear forms of the weak formulation, respectively. We solve the system \eqref{equ:weak form} numerically using finite elements methods. 

Finally, we form the Lagrangian objective by adding the constraints \eqref{equ:weak form}, for $m=1,...,M$, to the SAA of \eqref{equ:optimization functional}. The Lagrangian is then given as
\begin{align}
\mathcal{L}(\tau) =& \frac{1}{M} \sum_{m=1}^{M} \mathcal{Q}(\tau, \zeta_m) \nonumber\\ 
                   &+ \beta_V \frac{1}{M} \sum_{m=1}^{M} \mathcal{Q}^2(\tau, \zeta_m) - \Bar{\mathcal{Q}}^2 \nonumber \\
                   &+ \beta_P P(\tau) \nonumber\\
                   &+ \sum_{m=1}^{M} \left(a(u^\text{sca}_m, v_m;\tau, \zeta_m) - b(v_m)\right).
\label{equ:Lagrangian}
\end{align}
The test functions $v_m$ play the role of the Lagrange multipliers in the Lagrangian \eqref{equ:Lagrangian}.

We solve the optimization problem of minimizing $\mathcal{L}$, upon discretization, using BFGS method. To compute the gradient required for BFGS, we form an adjoint-based gradient \cite{gunzburger2002perspectives} which require $M$ forward problem, \eqref{equ:weak form}, solves and $M$ corresponding adjoint problem solves. Computing the adjoint-based gradient is scalable since the number of PDE solves required, $2M$, is independent of the number of DOFs of the discretization of $\tau$. We use the python library Stochastic Optimization under high-dimensional Uncertainty in Python (SOUPy) for the SAA based optimization \cite{Chen_Github}.

\section{Numerical results}\label{sec:PNJ Control Results}

We present deterministic PNJ design results in sections \ref{sec:PNJ Control Results:Optimal lens design}
and \ref{sec:PNJ Control Results:Steering the PNJ}. In section \ref{sec:PNJ Control Results:Effect of manufacturing noise on PNJ location and intensity}, we study the effect of the manufacturing error through forward uncertainty quantification. In section \ref{sec:PNJ Control Results:Controlling the PNJ under uncertainty}, we present application of SAA OUU to the PNJ design problem.

\subsection{Optimal lens design}\label{sec:PNJ Control Results:Optimal lens design}

\begin{figure}[!htb]
    \centering
    \includegraphics[width=3.2in]{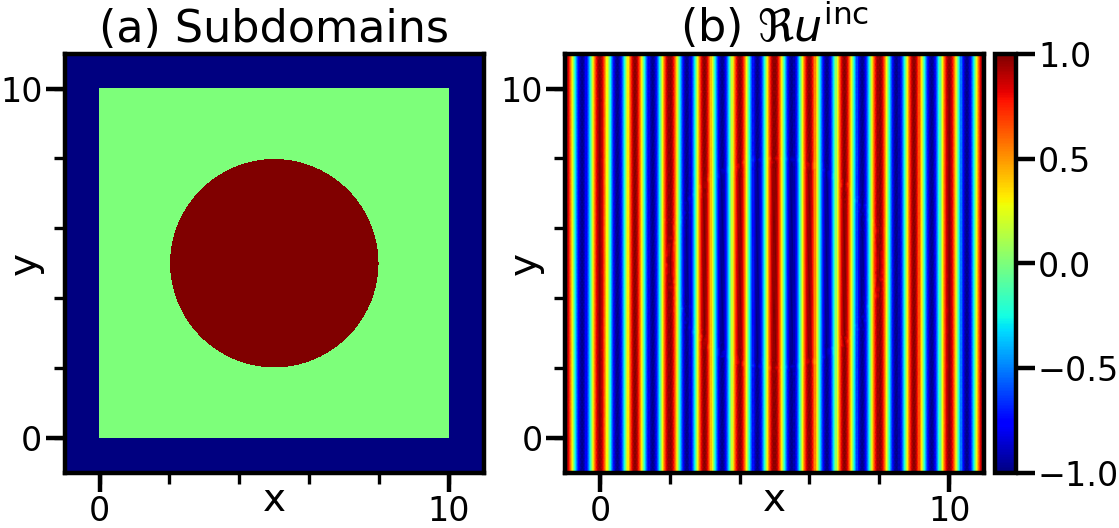}
    \caption{(a) Simulation subdomains: the lens $\mathcal{D}$ (dark red), the background medium (green), and the PML layer (dark blue). (b) The real part of the incident wave $u^\text{inc}$.}
    \label{fig:subdomain_ur_inc}
\end{figure}

For all the numerical experiments in this paper, we assume a circular lens of radius 3 units. We set the simulation domain to be of 10 by 10 units and the PML layer width to be 1 unit, Fig.~\ref{fig:subdomain_ur_inc}\emph{a}. The incident wave in our simulation is a plane wave given by  $u^{\rm inc}(\textbf{x})=\exp(ik_0\textbf{x}.\widehat{\textbf{b}})$. The real part of $u^\text{inc}$ is shown in Fig.~\ref{fig:subdomain_ur_inc}\emph{b}.

\begin{figure}[!htb]
    \centering
    \includegraphics[width=3.2in]{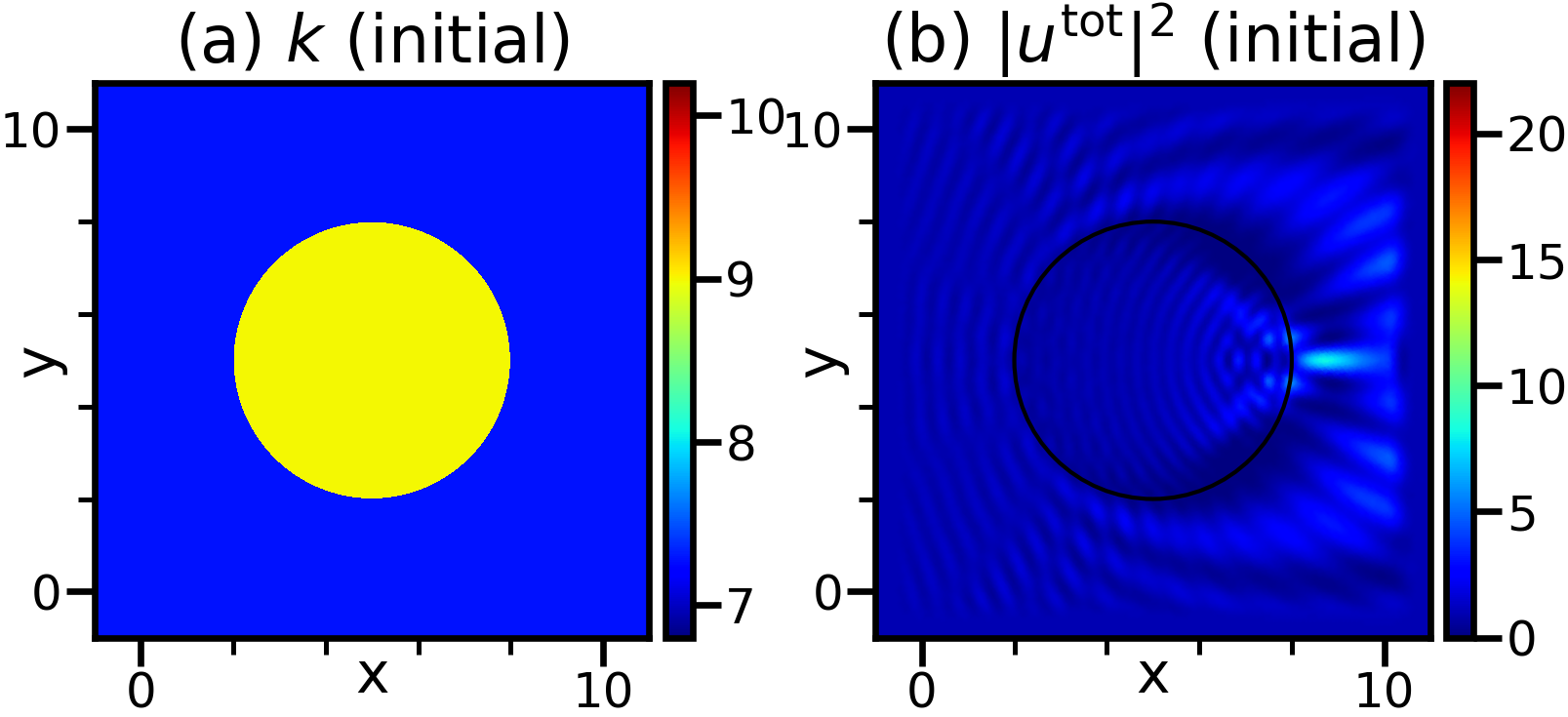}
    \caption{(a) Initial homogeneous lens profile. (b) The total amplitude profile squared that corresponds to the initial lens profile in (a).}
    \label{fig:deterministic initial}
\end{figure}

\begin{figure}[!htb]
    \centering
    \includegraphics[width=3.2in]{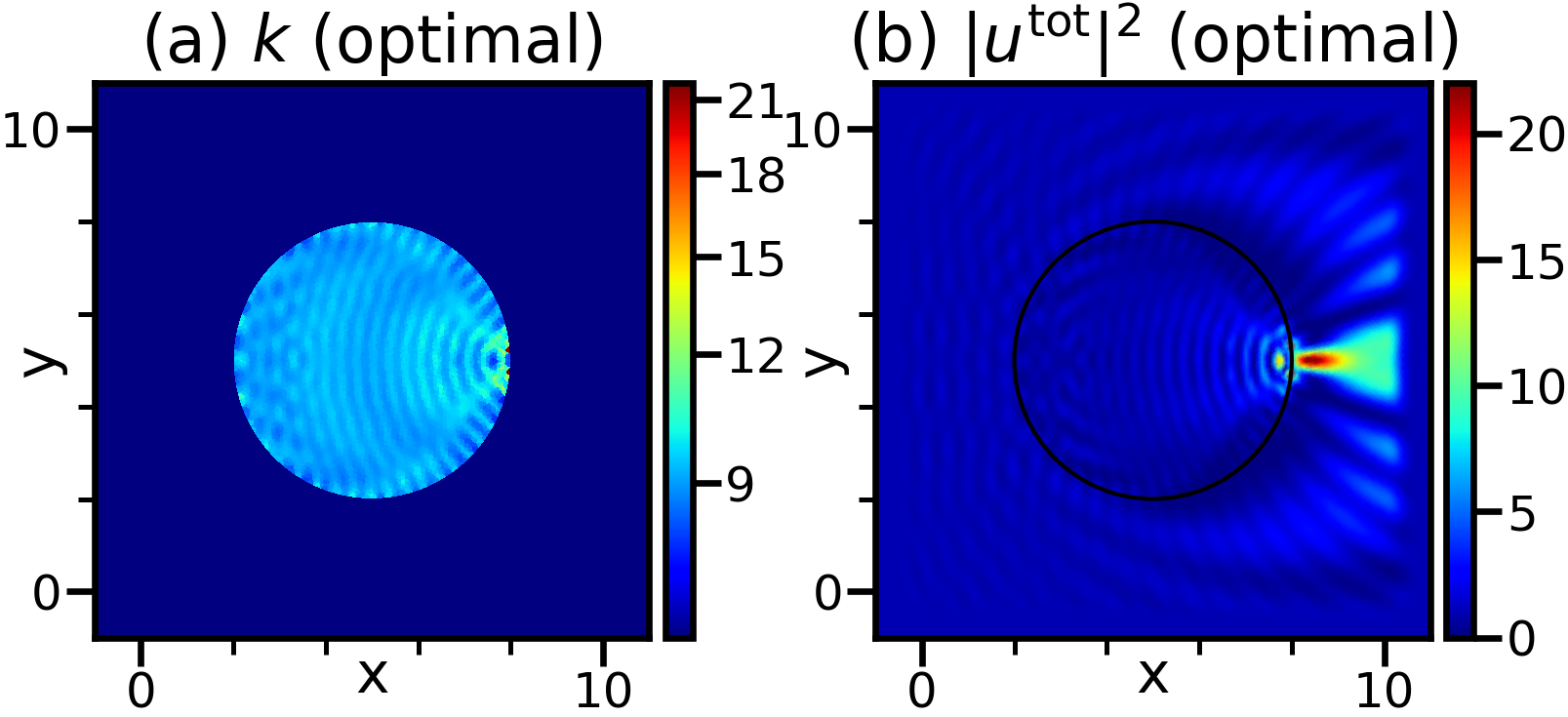}\\
    \includegraphics[width=1.6in]{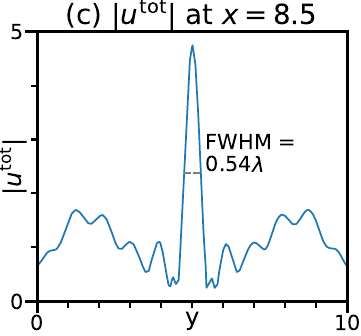}
    \caption{(a) Optimal heterogeneous lens profile  and (b) the corresponding total amplitude squared for the case in which $x_\text{PNJ} = (8.5,5)$ and $A_\text{PNJ}=20$. (c) A slice of total amplitude profile at $x=8.5$ (blue line). The extent of the full width half maximum (FWHM) is marked in a dashed black line.}
    \label{fig:deterministic optimal}
\end{figure}

For the initial homogeneous lens profile, Fig.~\ref{fig:deterministic initial}\emph{a}, the resulting total amplitude profile squared is shown in Fig.~\ref{fig:deterministic initial}\emph{b}. We note a PNJ-like formation with maximum amplitude square of about 8 units. Starting from this initial lens profile, we solve the optimization problem of minimizing the deterministic form of the objective \eqref{equ:Lagrangian} ($M=1$ and $\zeta_m=0$) where $x_\text{PNJ} = (8.5,5)$ and  $A_\text{PNJ}=20$. The solution, the optimal profile, is shown in  Fig.~\ref{fig:deterministic optimal}\emph{a}. The achieved total amplitude squared field that corresponds to the optimal profile is shown in Fig.~\ref{fig:deterministic optimal}\emph{b}. We observe that a PNJ-like formation in the desired location $x_\text{PNJ}$ is evident  and that the maximum amplitude square at the peak of the PNJ formation is significantly larger than the initial amplitude, $\sim22$ units. The full width half maximum (FWHM) of the formed PNJ is about $0.54\lambda$, Fig.~\ref{fig:deterministic optimal}\emph{c}.

\begin{figure}[!htb]
    \centering
    \includegraphics[width=3.2in]{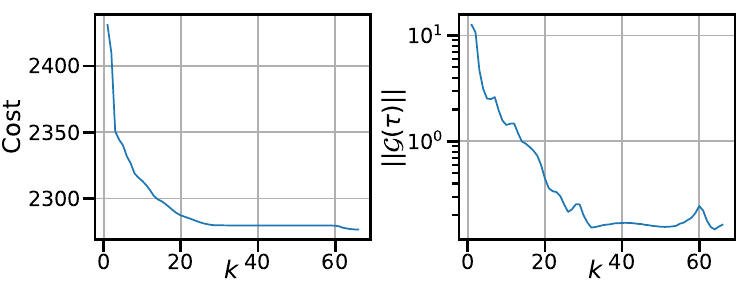}
    \caption{Convergence of BFGS method to the optimal solution in Fig.~\ref{fig:deterministic optimal}. (left) Cost value versus iteration number. (right) Gradient norm versus iteration number.}
    \label{fig:deterministic convergence}
\end{figure}

Reduction of the cost and the gradient of the optimization objective using  BFGS method \cite{byrd1995limited} is shown in Fig.~\ref{fig:deterministic optimal}\emph{b}. At around iteration $30$, both the cost functional and the gradient norm plateau around constant values as shown in Fig.~\ref{fig:deterministic convergence}. The optimization is terminated after reaching a maximum backtracking limit of $70$.  The gradient is reduced, significantly, approximately two orders of magnitude. The cost functional is still relatively large after termination because attaining the desired amplitude of $A_\text{PNJ}=20$ is not possible for the model that we specify, hence the large discrepancy between $A_\text{PNJ}$ and the achieved amplitude $\sim5$.

\subsection{Radial and Angular shift in PNJ location}\label{sec:PNJ Control Results:Steering the PNJ}

\begin{figure}[!htb]
    \centering
    \includegraphics[width=3.2in]{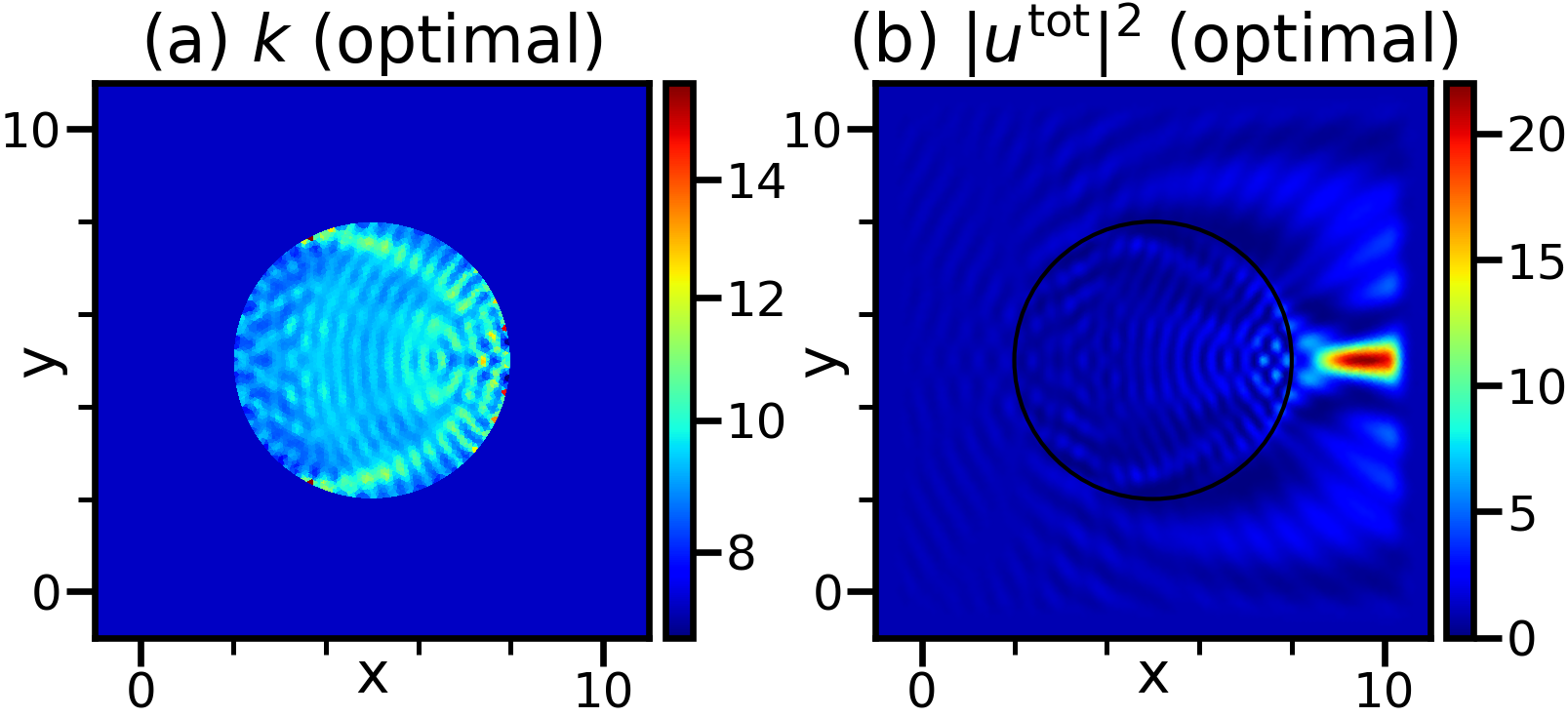}
    \caption{PNJ design with a radial shift in the desired PNJ location. (a) Optimal heterogeneous lens profile and (b) the corresponding total amplitude squared for the case in which $x_\text{PNJ} = (9.5,5)$ and $A_\text{PNJ}=20$.}
    \label{fig:steering radial}
\end{figure}

In this section we present two numerical experiments. In the first one, the desired location is shifted along the radius, $x_\text{PNJ} = (9.5,5)$ and  $A_\text{PNJ}=20$. We obtain an optimal lens design, Fig.~\ref{fig:steering radial}\emph{a}, that is considerably different from the one we obtain in Fig.~\ref{fig:deterministic optimal} and achieves the desired radial shift in the PNJ location. We obtain a PNJ that peaks at $x_\text{PNJ} = (9.5,5)$, Fig.~\ref{fig:steering radial}\emph{b}. 

\begin{figure}[!htb]
    \centering
    \includegraphics[width=3.2in]{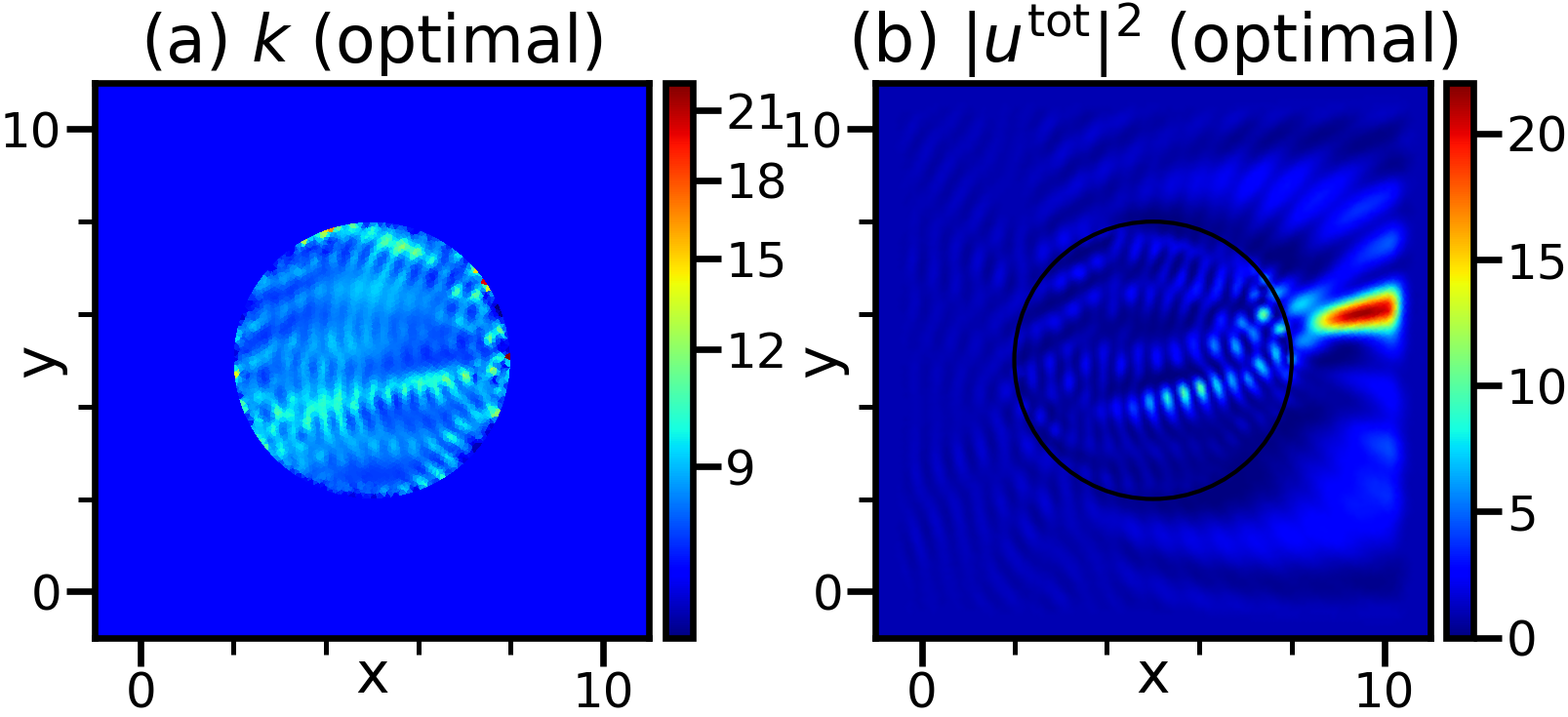}\\
    \includegraphics[width=3.2in]{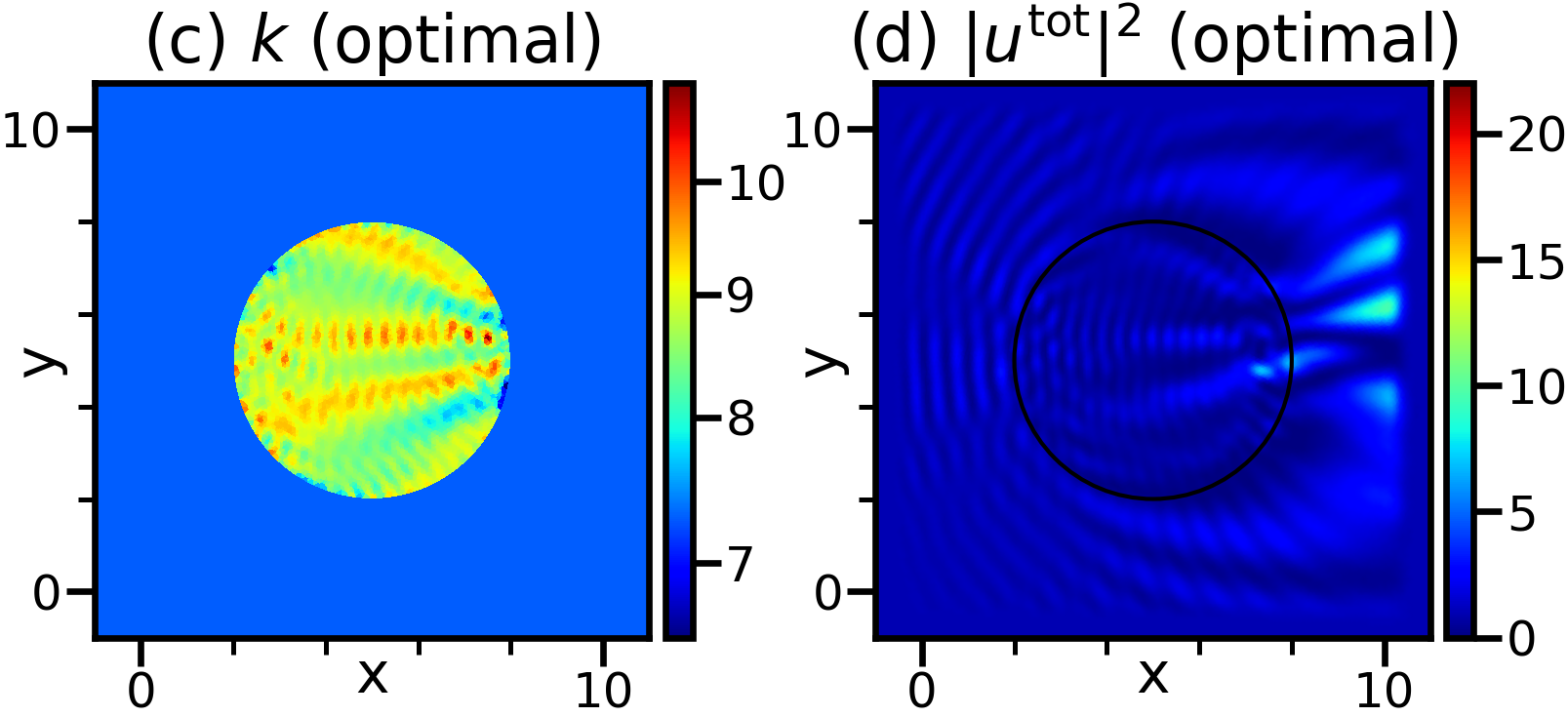}
    \caption{PNJ design with an angular shift in the desired PNJ location. (a) Optimal heterogeneous lens profile  and (b) the corresponding total amplitude squared for the case in which $x_\text{PNJ} = (8.5,6)$ and $A_\text{PNJ}=20$. (c) Optimal heterogeneous lens profile and (d) the corresponding total amplitude squared for the case in which $x_\text{PNJ} = (8.5,6)$ and $A_\text{PNJ}=7$.}
    \label{fig:steering angular}
\end{figure}

In the second numerical experiment, the desired location is shifted along the angular direction, $x_\text{PNJ} = (8.5,6)$ and  $A_\text{PNJ}=20$. In this case, the optimal lens design that we obtain, Fig.~\ref{fig:steering angular}\emph{a},  is non-symmetric and distinct from the previous lens designs. It effectively ``bends'' the light to achieve the desired PNJ that peaks at $x_\text{PNJ} = (8.5,6)$, Fig.~\ref{fig:steering angular}\emph{b}. 

Note that the desired PNJ amplitude  in the two scenarios we discuss here is $\sim5$ while  $A_\text{PNJ}=20$.  To illustrate the effect of choosing the value of $A_\text{PNJ}$, we present a scenario in which $5 < A_\text{PNJ}< 20$. We choose $A_\text{PNJ}=7$ and keep $x_\text{PNJ} = (8.5,6)$ unchanged. The optimization in this case fails to produce the PNJ phenomena and the intense light energy is scattered among multiple formations, Fig.~\ref{fig:steering angular}\emph{c}~and~\ref{fig:steering angular}\emph{d}.

\subsection{Effect of manufacturing noise on PNJ location and intensity}\label{sec:PNJ Control Results:Effect of manufacturing noise on PNJ location and intensity}

\begin{figure}[!htb]
    \centering
    \includegraphics[width=3.2in]{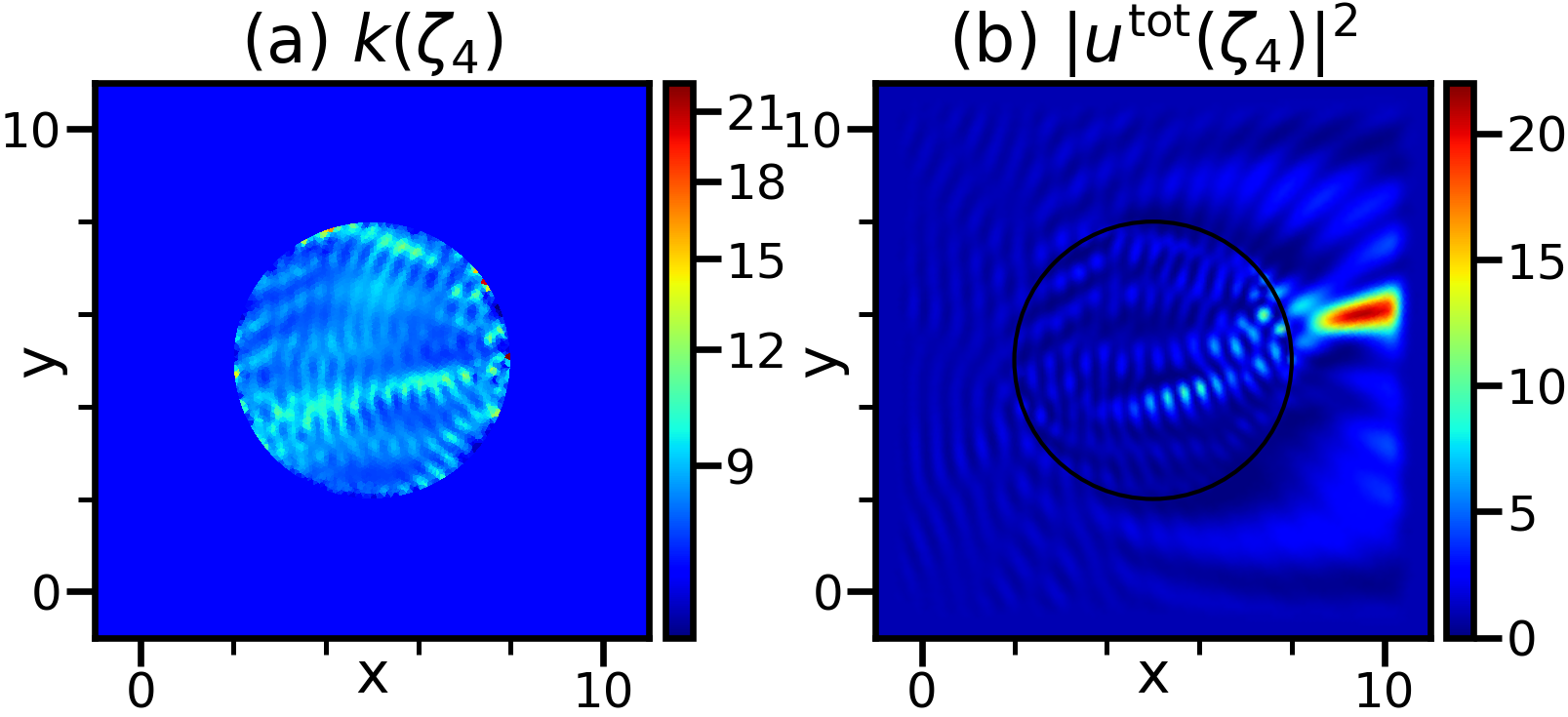}\\
    \includegraphics[width=3.2in]{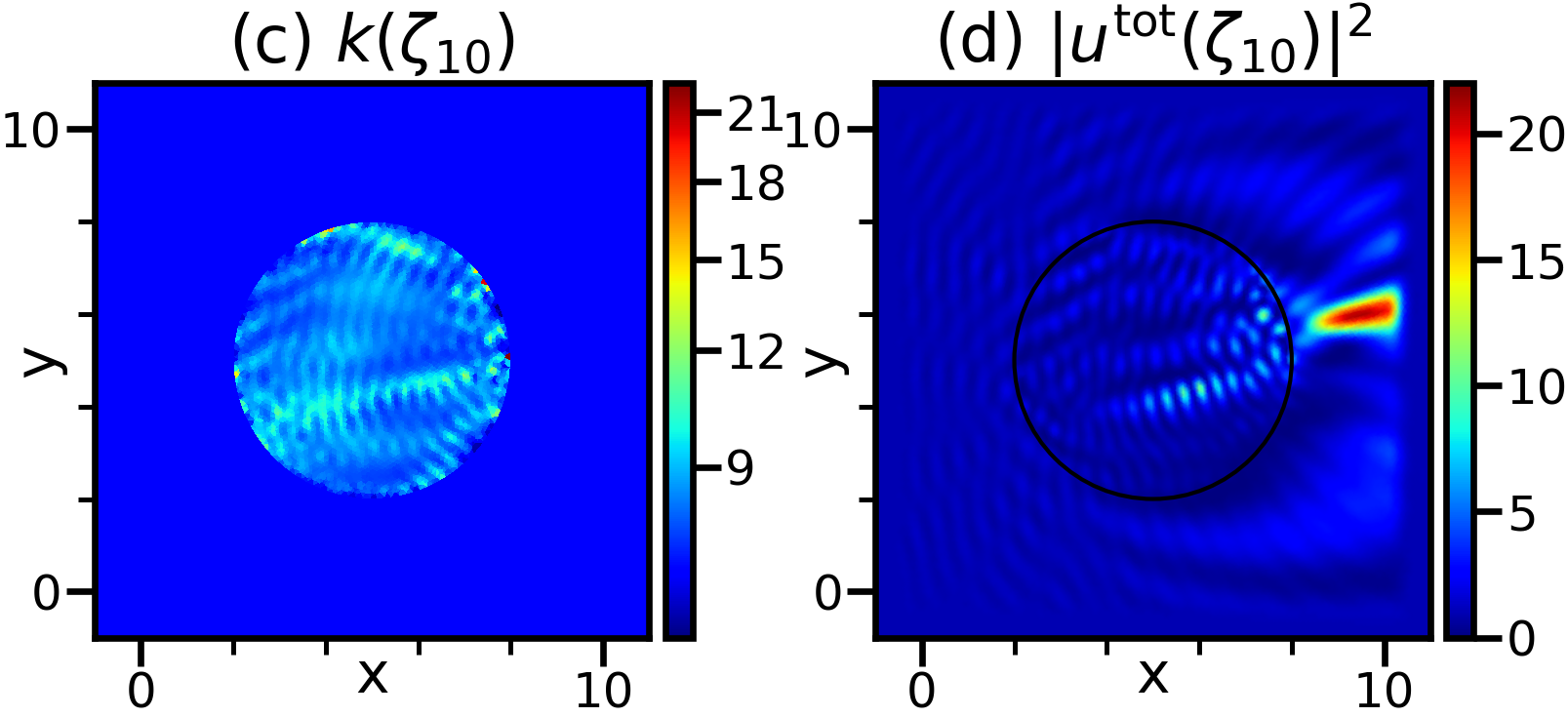}
    \caption{Effect of manufacturing errors on the result PNJ. (a) and (c) Two realizations of the optimal heterogeneous lens profile, Fig.~\ref{fig:steering angular}\emph{a}, polluted by two different realizations of the manufacturing noise. (b) and (d) The corresponding total amplitude squared. $x_\text{PNJ} = (8.5,6)$ and $A_\text{PNJ}=20$.}
    \label{fig:noise deterministic}
\end{figure}

By applying forward uncertainty quantification, we study the effect of manufacturing errors in the lens design on the formed PNJ. We add manufacturing error realizations, obtained from the same distribution from which the realizations in Fig.~\ref{fig:noise samples} are obtained, to the optimal lens in Fig.~\ref{fig:steering angular}\emph{a} and solve the forward problem \eqref{equ:helmholtz} to obtain the corresponding total amplitude squared. We repeat this process for $M=15$ error realizations. In Fig.~\ref{fig:noise deterministic}, we show two cases of the 15 cases. The location and intensity of the result PNJ varies based on the noise realization, e.g. the PNJ in Fig.~\ref{fig:noise deterministic}\emph{b} is slightly different than the one in Fig.~\ref{fig:noise deterministic}\emph{d}. This shows that the presence of manufacturing noise of this magnitude affect the properties of the desired PNJ. 

\begin{figure}[!htb]
    \centering
    \includegraphics[width=3.2in]{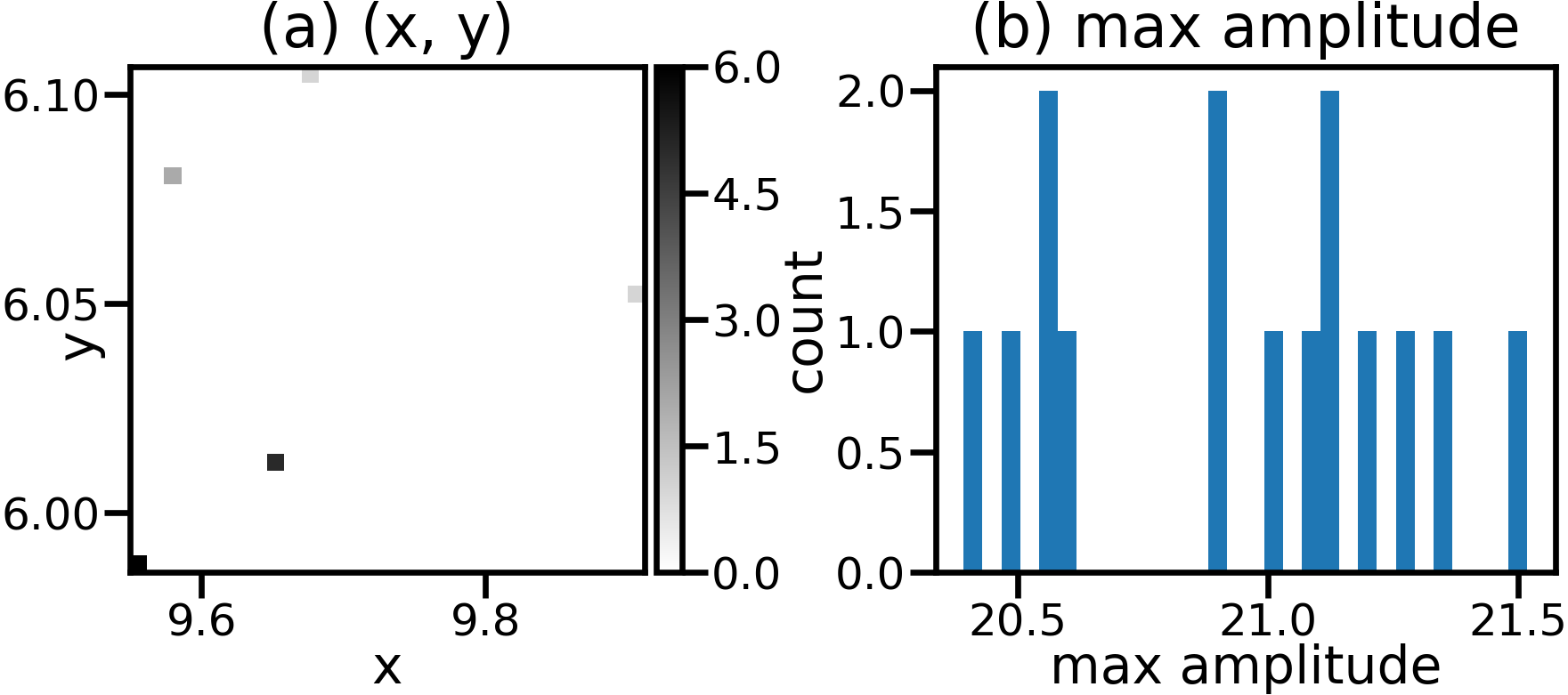}
    \caption{Histogram of the PNJ features resulting from 15 lens profiles. Each of those profiles is the deterministic-based optimal lens, Fig.~\ref{fig:steering angular}\emph{a}, polluted by a different manufacturing error realization. (a) PNJ location histogram. (b) PNJ maximum amplitude squared histogram.}
    \label{fig:noise deterministic histo}
\end{figure}

To summarize the variation of PNJ properties due to the $M=15$ error realizations, we plot a histogram of the PNJ maximum intensity location and maximum amplitude squared, Fig.~\ref{fig:noise deterministic histo}. We note that the maximum amplitude is attained at five different locations with varying statistical frequency, Fig.~\ref{fig:noise deterministic histo}\emph{a}. These locations are discrete because they correspond to finite element mesh vertices. The maximum total amplitude squared also varies. It attains values ranging from $\sim20.4$ to $\sim21.5$, Fig.~\ref{fig:noise deterministic histo}\emph{b}.

\subsection{Designing the PNJ under uncertainty}\label{sec:PNJ Control Results:Controlling the PNJ under uncertainty}

\begin{figure}[!htb]
    \centering
    \includegraphics[width=3.2in]{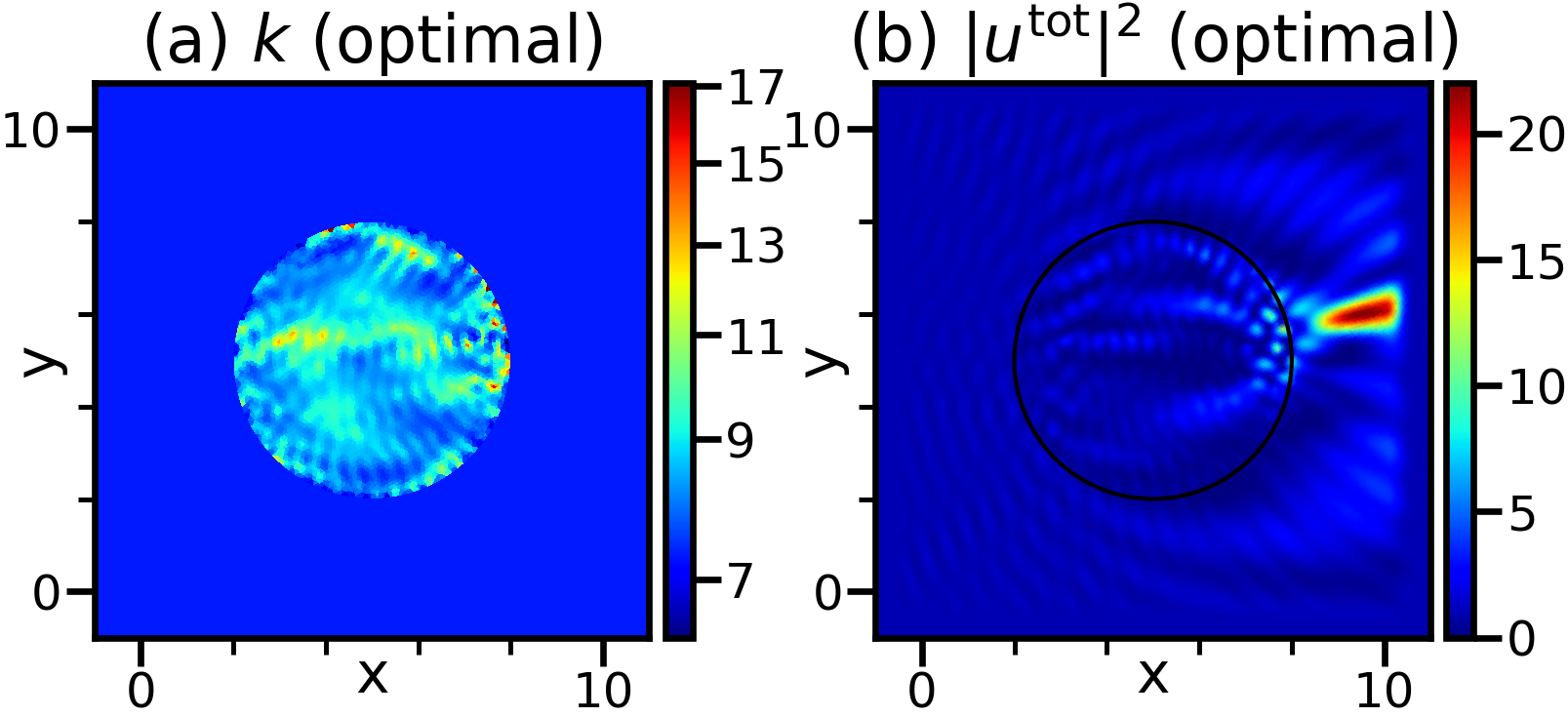}
    \caption{PNJ design under manufacturing uncertainty. (a) The optimal heterogeneous lens profile as a result of minimizing \eqref{equ:Lagrangian}. (b) The corresponding total amplitude squared. $x_\text{PNJ} = (8.5,6)$ and $A_\text{PNJ}=20$.}
    \label{fig:noise SAA}
\end{figure}

Here we solve the problem of finding the optimal lens design while taking the manufacturing errors into account via using the SAA based OUU formulation as described in sections \ref{sec:The Control Problem: OUU} and \ref{sec:The Control Problem: numerical methods}. We obtain an optimal lens profile, Fig.~\ref{fig:noise SAA}\emph{a}, that has visibly different features form the lens profile obtained in the corresponding deterministic setup, Fig.~\ref{fig:steering angular}. The PNJ formation in the desired location, $x_\text{PNJ} = (8.5,6)$, is evident in the corresponding total amplitude squared profile, Fig.~\ref{fig:noise SAA}\emph{b}. In this experiment, $M=15$ and $A_\text{PNJ}=20$. 

\begin{figure}[!htb]
    \centering
    \includegraphics[width=3.2in]{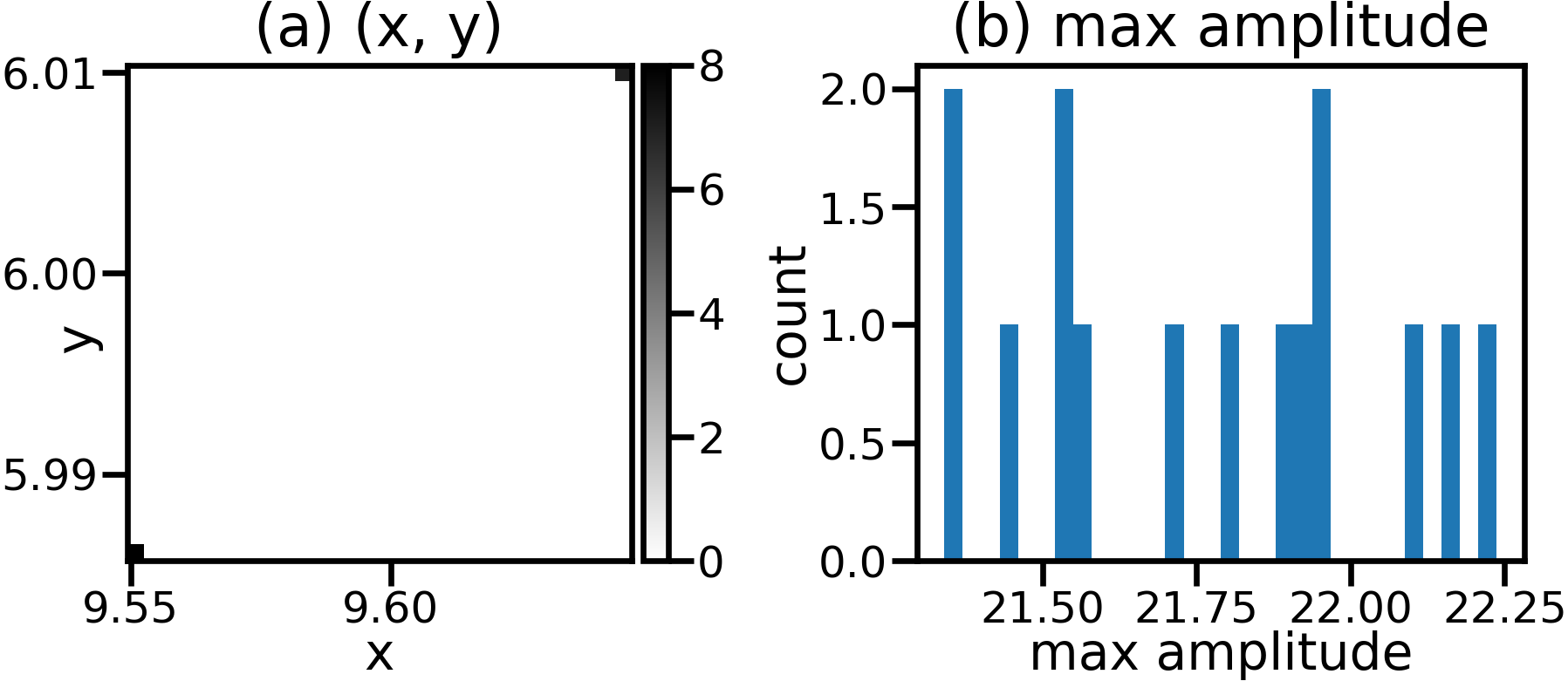}
    \caption{Histogram of the PNJ features resulting from 15 lens profiles. Each of those profiles is the SAA-based optimal lens, Fig.~\ref{fig:noise SAA}\emph{a}, polluted by a different manufacturing error realization. (a) PNJ location histogram. (b) PNJ maximum amplitude squared histogram.}
    \label{fig:noise SAA histo}
\end{figure}

Analogous to the deterministic case, we summarize the PNJ properties for this SAA OUU case in histograms, Fig.~\ref{fig:noise SAA histo}, to study the effect of the same 15 manufacturing error realizations on the produced PNJ. We see that using SAA OUU approach leads to more robust PNJ properties at the presence of manufacturing errors: For the 15 error samples, the attained PNJ locations are realized in only two points in space (compared to five points in Fig.~\ref{fig:noise deterministic histo}\emph{a}) and the attained maximum amplitudes are generally higher than those in the deterministic case (Fig.~\ref{fig:noise deterministic histo}\emph{b}). 

In table \ref{tab:tab1}\emph{a}, we provide the mean and the variance for each of the three quantities, the x and y coordinates of the PNJ peak and the total amplitude square value at the peak, for both the deterministic and the SAA OUU cases. In the SAA OUU case, the variance of the PNJ location is notably smaller and the mean of the maximum amplitude is larger when compared to the deterministic case.  In table \ref{tab:tab1}\emph{b}, we show the same quantities for a slightly different numerical experiment in which the noise amplitude is doubled. The results in this case are an evident confirmation of the advantage of using the SAA OUU formulation. The variances of the three quantities are reduced when using SAA OUU compared to the deterministic optimization, and the mean of the attained maximum amplitude is larger.

\begin{table}[!htb]
\centering
(a) $\delta=25$, $\gamma=2.5$\\
\begin{tabular}{ |l|c|c|c| } 
\hline
 & $x$ & $y$ & maximum  \\
  &  &  &   amplitude \\
\hline
mean (deterministic) & 9.61 & 6.01 & 20.93 \\ 
variance (deterministic) & 0.0083 & 0.0015 & 0.112\\ 
\hline
mean (SAA) & 9.59 & 5.99 & 21.76 \\ 
variance (SAA) & 0.0023 & 0.00015 & 0.078\\ 
\hline
\end{tabular}

\centering
\vspace{15pt}
(b) $\delta=12.5$, $\gamma=1.25$\\
\begin{tabular}{ |l|c|c|c| } 
\hline
 & $x$ & $y$ & maximum  \\
  &  &  &   amplitude \\
\hline
mean (deterministic) & 9.6 & 6.02 & 19.4 \\ 
variance (deterministic) & 0.02 & 0.005 & 1.27\\ 
\hline
mean (SAA) & 9.61 & 6.01 & 21.76 \\ 
variance (SAA) & 0.008 & 0.0017 & 0.33\\ 
\hline
\end{tabular}
\caption{Mean and variance values of PNJ properties: location $(x,y)$ and maximum amplitude, resulting from noise polluted lenses. (Table (a), first and second rows)  the statistics are computed from the optimal lens in the deterministic case, Fig.~\ref{fig:steering angular}\emph{a}, by polluting it by 15 different manufacturing error realizations. (Table (a), third and fourth rows)  the statistics are computed from the optimal lens in the SAA OUU case, Fig.~\ref{fig:noise SAA}\emph{a}, by polluting it by the same 15 manufacturing error realizations. For both cases summarized in Table (a), $\delta=25$ and $\gamma=2.5$. Table (b) reports the statistics of a similar study that differs only in the manufacturing error model where $\delta=12.5$ and $\gamma=1.25$, leading to a doubled error level compared to the case in Table (a).
}
\label{tab:tab1}
\end{table}

\begin{figure}[!htb]
    \centering
    \includegraphics[width=3.2in]{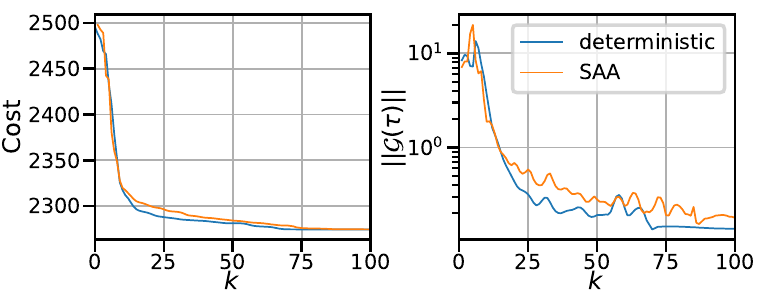}
    \caption{Convergence of BFGS method to the deterministic optimal solution in Fig.~\ref{fig:steering angular} (in blue) and to the SAA-based optimal solution in Fig.~\ref{fig:noise SAA} (in orange). (left) Cost value versus iteration number. (right) Gradient norm versus iteration number.}
    \label{fig:convergence SAA determinisitc}
\end{figure}

In Fig.~\ref{fig:convergence SAA determinisitc}, we compare the convergence results of the deterministic and the SAA runs. Overall, the convergence patterns are similar for the two cases. However, if we take a look at the gradient norm plot, Fig.~\ref{fig:convergence SAA determinisitc} (right), we can see that the convergence of the SAA run is slightly slower compared to the deterministic run. This is expected because of the complicated SAA objective function that has $M$ cost functions to be minimized simultaneously.

\section{Conclusion}
In this work, we demonstrate that designing PNJs through an optimized heterogeneous lens profile design is computationally feasible and provides flexibility in designing the location of the desired PNJ. Additionally, incorporating manufacturing uncertainty in the optimization problem of the heterogeneous lens design gives a non-trivial optimal lens design that achieves robust PNJ design. Some future work will be on alternative stochastic formulation using probability or chance constraint \cite{chen2021taylor} to control the defect rate, optimal design with design constraint to facilitate more feasible manufacturing, and experimental validation of the proposed computational results.

\section*{Acknowledgments} 
Karamehmedovi\'c and Alghamdi were supported by The Villum Foundation (grant no. 25893). Karamehmedovi\'c also received funding within the project (20FUN02/f10 POLight) from the EMPIR programme cofinanced by the Participating States and by the European Union’s Horizon 2020 research and innovation programme. Chen was partially supported by NSF DMS grant 2012453, USA.

\appendix

\section*{Appendix: Matern Covariance}
\label{app: appendixA}
\setcounter{equation}{0}

We assume $\zeta \sim \mathcal{N}(0,\mathcal{C})$ where $\mathcal{N}(0,\mathcal{C})$ is a Mat\'ern class infinite dimensional Gaussian random field with zero mean and  covariance operator $\mathcal{C}$. To sample from $\mathcal{N}(0,\mathcal{C})$ efficiently, we employ the link between Mat\'ern class Gaussian random fields and elliptic Stochastic Partial Differential Equations (SPDE) \cite{lindgren2010explicit,VillaPetraGhattas21}.
\begin{align}
(\delta I- \gamma \Delta)^{\alpha/2}\zeta(x) &= w(x) \quad\text{in}\; \mathcal{D}\nonumber \\
\nabla \zeta \cdot n &= 0 \quad\text{on}\;\delta \mathcal{D},\nonumber
\label{SPDE}
\end{align}
where $w(x)$ is white noise. The choice of $\delta$ and $\gamma$ dictates the variance and the correlation length of the Gaussian field spacial features. $\alpha$ is a smoothing parameter that satisfies  $\alpha > d/2$ for variance boundedness. For the choice of  $\alpha = 2$, generating a sample from this field, amounts to solving the elliptic system above, with a white noise sample $w(x)$ as the right-hand side, and a matrix-vector product \cite{lindgren2010explicit}. Solving this system can be done scalably using multigrid method \cite{VillaPetraGhattas21}.

\bibliographystyle{unsrtnat}
\bibliography{references}

\end{document}